\documentclass[letterpaper]{article} 
\usepackage{aaai25} 
\usepackage{times}  
\usepackage{helvet}  
\usepackage{courier}  
\usepackage[hyphens]{url}  
\usepackage{graphicx} 
\usepackage{natbib}  
\usepackage{caption} 
\usepackage{comment}
\usepackage{algorithm}
\usepackage{algorithmic}

\usepackage{newfloat}
\usepackage{listings}
\DeclareCaptionStyle{ruled}{labelfont=normalfont,labelsep=colon,strut=off} 
\floatstyle{ruled}
\newfloat{listing}{tb}{lst}{}
\floatname{listing}{Listing}
\title{VEL: A Formally Verified Reasoner for OWL2 EL Profile}
\author {
    Atalay Mert Ileri\textsuperscript{\rm 1},
    Nalen Rangarajan\textsuperscript{\rm 1},
    Jack Cannell\textsuperscript{\rm 1},
    Hande McGinty\textsuperscript{\rm 1}
}
\affiliations {
    \textsuperscript{\rm 1}Kansas State University, USA\\
    atalay@ksu.edu, 
    nalenrangarajan@ksu.edu, jdcannell@ksu.edu,
    hande@ksu.edu
}

\usepackage{todonotes}
\usepackage{amssymb}
\usepackage{amsmath}
\usepackage{comment}
\newtheorem{theorem}{Theorem}

\newtheorem{lemma}[theorem]{Lemma}
\newcommand{\Nt}{\{t\}}
\newcommand{\baa}{Baader et. al. }

\newcommand{\ci}{\sqsubseteq}
\newcommand{\ifff}{\leftrightarrow}
\newcommand{\crule}[1]{\textbf{CR#1}}
\newcommand{\cbox}{\mathcal{C}}
\newcommand{\model}{\mathcal{I}}
\newcommand{\BC}{\mathbf{BC}}
\newcommand{\BCm}{\mathbf{BC}^-}
\newcommand{\reach}{\rightsquigarrow}
\newcommand{\ELpp}{$\mathcal{EL}{++}$}

\begin{document}

\maketitle

\begin{abstract}
    Over the past two decades, the Web Ontology Language (OWL) has been instrumental in advancing the development of ontologies and knowledge graphs, providing a structured framework that enhances the semantic integration of data. However, the reliability of deductive reasoning within these systems remains challenging, as evidenced by inconsistencies among popular reasoners in recent competitions. This evidence underscores the limitations of current testing-based methodologies, particularly in high-stakes domains such as healthcare. To mitigate these issues, in this paper, we have developed VEL, a formally verified \ELpp reasoner equipped with machine-checkable correctness proofs that ensure the validity of outputs across all possible inputs. This formalization, based on Baader et al.'s algorithm \cite{Pushing-the-EL-Envelope}, has been transformed into executable OCaml code using the Coq proof assistant's extraction capabilities. Our formalization revealed several errors in the original completeness proofs, which led to changes to the algorithm to ensure its completeness. Our work demonstrates the necessity of mechanization of reasoning algorithms to ensure their correctness at theoretical and implementation levels.
\end{abstract}

%

\section{Introduction}
During the nearly three decades since its initial proposal, Semantic Web(SW) has become a mature technology, which can support and enhance existing applications in several scenarios, ranging from e-commerce \cite{necula2018enhancement} to Human-Device interaction \cite{perera2020personalised}, and product configuration \cite{bischof2018integrating}, just to name a few. The fundamental concept of the Semantic Web is to provide a semantic layer for data, enabling systems to manipulate this data through chains of meaningful deduction steps based on sound rules. This semantic framework allows SW data processing to offer meaningful explanations, which contrasts with the often opaque nature of sub-symbolic technologies like neural networks. However, a closer examination of SW systems reveals that the sophistication of their reasoning abilities can vary significantly \cite{colucci2024review}.
Furthermore, due to the increased popularity of Neurosymbolic AI in recent years \cite{nai}, there is a renewed interest in deductive reasoning algorithms.

Due to their important role, it is necessary to ensure that the reasoner results are correct. However, like any other software system, reasoner implementations are susceptible to software bugs that undermine their correctness. A recent competition for reasoners showed that some of the widely used reasoners contain bugs since the results of the different reasoners didn't agree on the same queries.~\cite{The-OWL-Reasoner-Evaluation-ORE-2015-Competition-Report} The ground truth needed to be determined by the majority vote. Reasoners' inability to agree on a correct result shows that testing-based methodologies cannot provide strong correctness guarantees required for critical domains such as healthcare. 

We developed VEL, a formally verified \ELpp reasoner with machine-checkable correctness proofs to address this gap. Our proofs ensure the correctness of the output for each possible input. We based our formalization on Baader et al.'s algorithm.~\cite{Pushing-the-EL-Envelope} During our mechanization, we discovered two errors in the published correctness proof that required changes to the algorithm. We obtained the executable code through the extraction functionality of the Coq proof assistant~\cite{coq}.

\paragraph{Contributions}
This paper has the following contributions:
\begin{itemize}
\item Mechanized formalization of \ELpp description logic.
\item A modified subsumption checking algorithm that is provably correct.
\item An implementation of a subsumption checker with string and rational number concrete domains for \ELpp with a mechanized proof of correctness.
\item An executable OCaml code extracted from the verified implementation.
\end{itemize}

\section{Related Work}
In response to community feedback and the evolving needs of ontology developers, OWL2 was released as a W3C recommendation in 2009. Apart from addressing acute problems with expressivity, a goal in the development of OWL2 was to provide a robust platform for future development. OWL2 extends the W3C OWL Web Ontology Language with a small but useful set of features that have been requested by users, for which effective reasoning algorithms can be built, and those OWL tool developers are willing to support. Considerable progress has been achieved in the development of tool support for OWL2. The new syntax is currently supported by the new version of the OWL API. The widely used Protégé~\cite{protege} system has been extended with support for the additional constructs provided by OWL2 and most recently gaining community support from volunteer groups such as OBO Foundry Ontologies \cite{obo}.

The standardization of OWL has sparked the development and adaption of a number of reasoners, including FacT++~\cite{fact++}, Pellet~\cite{pellet}, and HermiT~\cite{glimm2014hermit}, and ontology editors, including Protégé~\cite{protege}. However, mostly due to funding reasons and research interest shifting towards AI, the research for better reasoners was abandoned~\cite{knarm} despite the fact that recent work showed that more research is necessary for correct implementations~\cite{The-OWL-Reasoner-Evaluation-ORE-2015-Competition-Report}.

\subsubsection*{OWL Reasoners}

Over the past three decades, a number of RDF and OWL reasoners were implemented which are reviewed in various articles, most notably by Mishra et al.~\cite{mishra2011semantic} and Colucci et al.~\cite{colucci2024review}. However, for our purposes, we will list a few notable reasoners that have been dominating the field for the past two decades~\cite{knarm}.

\begin{itemize}
    \item \textbf{ELK:} ELK \cite{kazakov2012elk} is described as a high-performance reasoner for OWL EL ontologies. Unlike conventional tableau-based procedures, which test unknown subsumptions by trying to construct counter-models, the EL procedures derive new subsumptions explicitly using inference rules. 
    \item \textbf{FACT++ :} FaCT++ \cite{fact++} implements a tableaux decision procedure for the well-known $\mathcal{SHOIQ}$ description logic, with additional support for datatypes, including strings and integers. The system employs a wide range of performance-enhancing optimizations, including standard techniques (absorption and model merging) and newly developed ones (ordering heuristics and taxonomic classification).
    \item \textbf{HermiT:} HermiT \cite{glimm2014hermit} is the first publicly-available OWL reasoner based on a novel hypertableau calculus, which provides much more efficient reasoning than any previously known algorithm. 
    \item \textbf{Pellet:} Pellet \cite{pellet} is the first sound and complete OWL-DL reasoner with extensive support for reasoning with individuals (including nominal support and conjunctive query), user-defined datatypes, and debugging support for ontologies. 
    \item \textbf{Konclude:} Konclude \cite{steigmiller2014konclude} is a high-performance reasoner for the Description Logic $\mathcal{SROIQV}$. The supported ontology language is a superset of the logic underlying OWL 2 extended by nominal schemas, which allows for expressing arbitrary DL-safe rules.     
\end{itemize}

\subsubsection*{Mechanized Formal Logic}
A diverse selection of general-purpose logics such as propositional logic~\cite{A-Comprehensive-Formalization-of-Propositional-Logic}, first-order predicate logic~\cite{Formalizing-the-meta-theory-of-first-order-predicate-logic}, quantified modal logic~\cite{Towards-a-Coq-formalization-of-a-quantified-modal-logic}, and logic of bunched implications~\cite{Semantic-cut-elimination-for-the-logic-of-bunched-implications} are formalized in proof assistants for metatheoretical studies. Additionally, formalizations of more specialized logics such as separation logic~\cite{Iris-from-the-ground-up}, matching logic~\cite{Mechanizing-Matching-Logic-in-Coq}, alternating-time temporal logic~\cite{Alternating-Time-Temporal-Logic-in-the-Calculus-of-CoInductive-Constructions}, linear logic~\cite{Mechanizing-Focused-Linear-Logic-in-Coq}, and $\mathcal{ALC}$ description logic~\cite{Formally-Verified-Tableau-Based-Reasoners-for-a-Description-Logic} serve as tools for verifying software. Our work adds \ELpp description logic to this growing body of work.

\subsubsection*{Verified Reasoning}
Baader et al. use automated theorem proving to certify the results of the ELK reasoner for OWL2 EL ~\cite{First-Results-on-How-to-Certify-Subsumptions-Computed-by-the-EL-Reasoner-ELK-Using-the-Logical-Framework-with-Side-Conditions}. They implemented an algorithm that extracts a proof certificate from an ELK result and checks its correctness using LFSC proof checker. This approach incurs time and memory overheads due to certificate generation and proof checking at run-time and lacks support for required extensions such as concrete domains. In contrast, VEL does not incur any proof-checking overhead since the correctness proofs of our implementations are statically checked. Eliminating this overhead will make our reasoners more scalable compared to validation-based approaches. VEL also supports required extensions for wider applicability. 

Hidalgo-Doblado et al. formalized $\mathcal{ALC}$ description logic and implemented a formally verified tableau-based reasoner in PVS.~\cite{Formally-Verified-Tableau-Based-Reasoners-for-a-Description-Logic} Although it is a well-known description logic, $\mathcal{ALC}$ does not correspond to any OWL2 profile and lacks the support for datatypes and other extensions. These limitations restrict the usability of their implementation in applications. OWL2 guides VEL's development, and VEL supports necessary logic extensions to ensure' widespread usage of our implementations. 

\section{Formalization of \ELpp}
The first step in implementing a verified reasoner is formalizing the logic it will operate on. We decided to keep our formalization as generic as possible to facilitate its use in other contexts. In this section, we will explain each part of our formalization.

\subsection{Primitives}
We parameterized our formalization over primitives such as names and concrete domain elements. We defined the names as decidable infinite types. A decidable infinite type has two components: a decision procedure for checking the equality of two terms and a function that takes a list of terms and returns a term that is not in the list. Such parametrization allows using different types without changing the core implementation. As a result, our formalization is easier to integrate in different applications.

\subsubsection{Concrete Domains}
Our formalization models each domain as a record with four definitions and three proofs. Our definitions include the domain, which is a non-empty set of concrete domain elements, a set of predicate names with a decidable membership, the arities of each predicate as a partial function from predicate names to natural numbers, and an application function that determines the result of a predicate application to a list of concrete domain elements

Three required proofs are the specifications for arity definitions and the application function. The first one requires that arities are defined only for the predicates of the concrete domain. The second one provides a specification for the application function regarding input value validity. It states that if the function returns true for some predicate and list of concrete domain elements, then the predicate should belong to that concrete domain, each element should be in the domain of that concrete domain, and the number of elements should match the arity of the predicate.

During our formalization, we discovered that predicate names of different domains should also be distinct, which wasn't required in the original paper. This necessity arises because rule \crule{9} in classification needs to decide that two predicate names belong to different domains \cite{Pushing-the-EL-Envelope}. However, this is not a limitation in practice because shared names between domains can be replaced to ensure disjointness.

Some rules in the classification algorithm require checking the satisfiability of a conjunction of predicate expressions or the implication between predicate expressions, interpreted as a first-order formula. To be able to ensure the correctness of satisfiability and implication checker implementations, we defined an evaluation function that takes a first-order sentence composed of predicate expressions and an assignment to feature names and returns a boolean that represents if the given sentence is satisfiable under the given assignment. Then, we defined the satisfiability and implication checker specifications in terms of the results of the evaluations. We decided to use dependent typing to ensure the implementation of the functions to satisfy their specification.

\subsubsection{Language}
We defined the language as an inductive type where each constructor corresponds to an \ELpp concept description constructor. We also defined a well-formedness predicate for concrete domain descriptions to ensure that predicate names belong to one of the concrete domains and the number of feature names passed to the predicate matches its arity. Unlike other restrictions, we decided to define this as a predicate instead of using dependent typing to keep the equality of concept descriptions decidable.

On top of concept descriptions, we defined constraints and knowledge bases. Our knowledge bases consist of lists of names that can be used, a CBox, and the proof that the only names appearing in the CBox are the ones in the lists.

\subsubsection{Semantics}
We formalized the provided model-theoretic semantics for \ELpp. We first defined what we call ``a base interpretation''. A base interpretation maps each name to its corresponding structure in the model, namely, each individual name to an individual in the domain, each concept name to a set of individuals, each role name to a pair of individuals, and each feature name to a partial function from individuals to concrete domain elements.

Then, we defined the interpretation function that recursively constructs the interpretation of a concept description for a given base interpretation. Finally, we defined the notions of model and subsumption as described in \baa.

\section {Implementation}
\begin{figure}
    \centering
    \includegraphics[width=\linewidth]{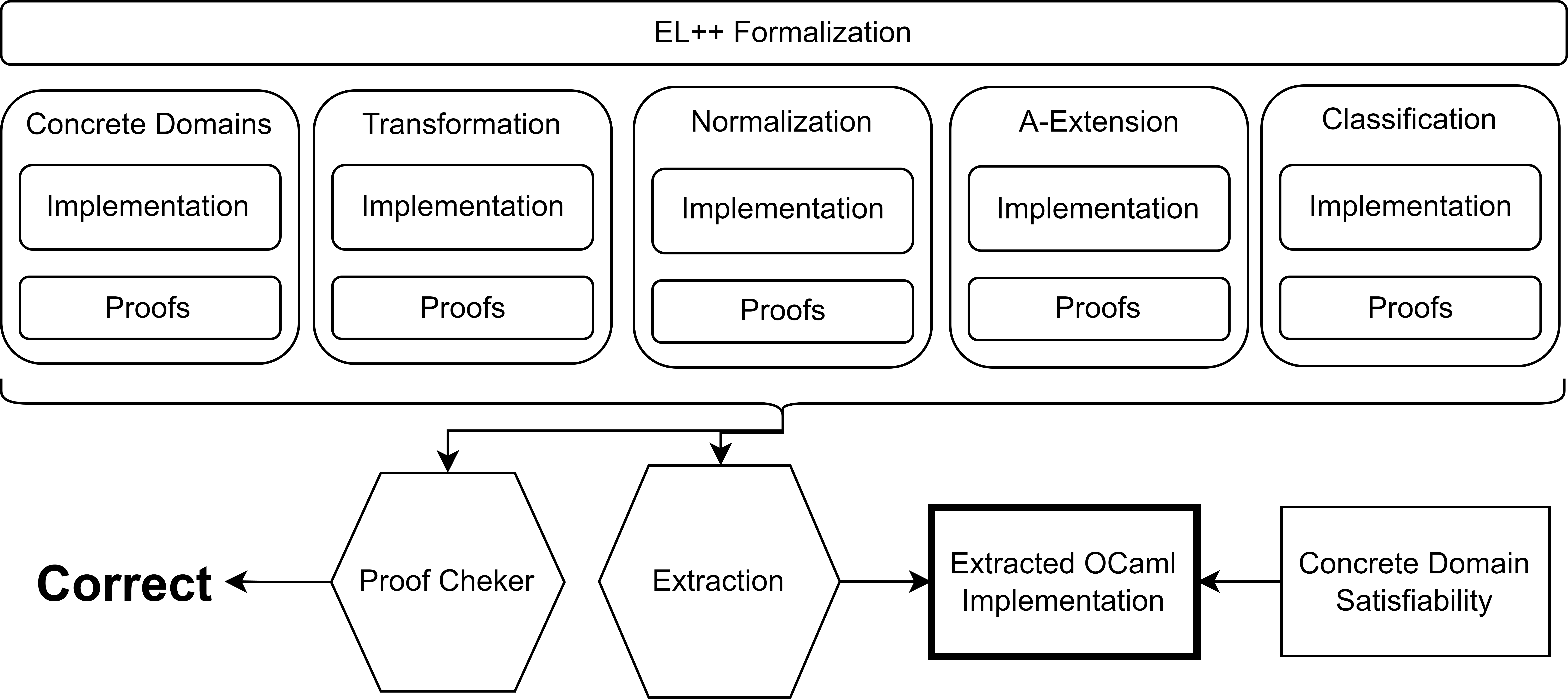}
    \caption{Structure of VEL}
    \label{project-structure}
\end{figure}

Our implementation of the reasoner consists of four parts: transformation, normalization, A-extension, and classification. 
Transformation changes the problem from checking the subsumption of arbitrary concept descriptions to checking the subsumption of two concept names.
Normalization breaks down complex constraints into simpler ones to make subsumption checking tractable. A-extension adds an axiom that ensures the completeness of the reasoner. Finally, classification computes concept inclusions from the constraints.
On top of our reasoner implementation, we implemented two concrete domains: rational numbers and strings. Project structure can be seen in Figure \ref{project-structure}. Rounded rectangles represent our components implemented in the proof assistant, hexagons represent the proof assistant's functionalities, and normal rectangles represent runnable code.

\subsection{Transformation}
The first step of checking subsumption $C \ci_\cbox D$ is transforming the knowledge base to add $A \ci C$ and $D \ci B$ where $A$ and $B$ are fresh concept names, as described in \baa This transformation changes the problem from subsumption checking of arbitrary concept descriptions to subsumption checking of basic concepts while preserving the subsumption relationship. It returns the transformed knowledge base and the newly generated names to be used in A-extension and final subsumption checking.

\subsection{Normalization}
We implemented the normalization in two steps, as presented in the paper. Both steps follow a similar pattern where each rule is repeatedly applied until none of the rules can be applied anymore. Again, we used dependent typing to ensure the normalization implementation satisfies its specifications.

\subsubsection{Formalization of the Rules}
We formalized each rule in two parts: the condition and the application. Such separation allowed us to use a higher-order selection function that removes and returns an eligible constraint given a rule condition or tells us no such constraint exists. Then, the chosen constraint is passed into the application function, which creates the new constraints and adds them to the CBox. 

This modular implementation allowed us to encapsulate each rule's effect on the CBox and prove the necessary theorems more easily, making the process tractable.

\subsubsection{Termination Measure} We implement each normalization part as a recursive function that finds candidates and applies the corresponding rules.
Defining a recursive function in Coq requires proving that it will terminate for all inputs. One way to do that is by providing a measure and proving that it decreases after each recursive call. We defined a measure based on the number of rule applications left.
However, naive counting wasn't enough to give us a decreasing measure since applying a rule could create new candidates for application. To circumvent that problem, we used a weighted count where applying a rule that can create new candidates decreases the measure more than the new candidates increase.
Listing \ref{measure-example} illustrates the part of the measure function for \textbf{NF3} rule that counts nested $\exists R.C$ structures in a concept description.

\subsubsection{Modified NF2}
We had to make one modification regarding \textbf{NF2} rule to the normalization routine to ensure that our measure decreases after every iteration. Measure for \textbf{NF2} counts the number of $\sqcap$'s that connect at least one complex concept description inside each constraint's left-hand side. This measure doesn't decrease when \textbf{NF2} is applied to a constraint $C \sqcap D \ci E$ where both $C$ and $D$ are complex concept descriptions. The resulting two constraints $A \sqcap D \ci E, C \ci A$ have the same total count with $C \sqcap D \ci E$. However, if we immediately apply \textbf{NF2} again to $A \sqcap D \ci E$, we get $A \sqcap B \ci E, D \ci B, C \ci A$, which has one less number of applications since both $A$ and $B$ are simple concept descriptions. So, we circumvented this problem by replacing \textbf{NF2} with a new rule that applies \textbf{NF2} twice when both concept descriptions are complex.

\begin{listing}
\caption{A part of the measure for \textbf{NF3} rule.}
\label{measure-example}
\begin{lstlisting}
Fixpoint count_NF3_CD(C) :=
match C with
| $X \sqcap Y \Rightarrow$  
    count_NF3_CD$(X)\ +\ $count_NF3_CD$(Y)$
| $\exists R. X \Rightarrow$  
    if $X \in \BC$ then $0$
    else count_NF3_CD$(X) + 3$
| _ $\Rightarrow 0$
end.
\end{lstlisting}
\end{listing}

\subsection{A-extension}
We added a preprocessing step to address the original completeness proof error. We call this step \textit{A-extension} of the knowledge base. Given a concept name $A$, 
A-extension of a CBox $\cbox$ is defined as

$$\cbox^{A+} := \cbox \cup \left\{ \Nt \ci \exists r_t.A \right\}$$

where $t$ and $r_t$ are fresh individual and role names, respectively. A-extension of a CBox ensures that in every $\cbox^{A+}$ model, $A^\model$ is nonempty. Nonemptiness plays a crucial role in the completeness proof, which will be explained in a future section.

We also proved that A-extension preserves subsumption with respect to $A$. This is due to the fact that each model of $\cbox$ where $A^\model$ is nonempty can be extended to obtain a model of $\cbox^{A+}$. Conversely, every model of $\cbox^{A+}$ is a model of $\cbox$. Therefore, the following theorem holds:

\begin{theorem}
$\forall\, \cbox, B \in N_C. A \ci_\cbox B \ifff A \ci_{\cbox^{A+}} B$
\end{theorem}

\subsection{Classification}
We followed a similar methodology to normalization by splitting rules into conditions and applications and then using a higher-level function to find candidates. This allowed us to use the same proof patterns we used in normalization. 

\paragraph{Reachability}
A crucial operation for classification is reachability. Reachability states that a sequence of concept descriptions exists in which each consecutive pair is related by a role between two concept descriptions. We implemented reachability as a depth-first search that starts from the end and constructs the path until it reaches the beginning. To verify it, we defined a predicate that implements the rules in the paper, then dependently typed the function so it returns \texttt{true} if and only if the predicate holds.

\paragraph{Classification Data Structure}
Our implementation of classification defines the two components, namely $S$ and $R$, as functions from concept descriptions to a list of concept descriptions and a list of pairs of concept descriptions, respectively. This definition simplifies the proof scripts but incurs a large run-time overhead. The overhead stems from the fact that each update adds a conditional at the top level of the function. Therefore, every lookup must traverse thousands of nested conditionals to compute the function. Replacement of classification implementation with a more efficient data structure left as future work.

\subsubsection{Termination Measure}
For all the rules except \crule{6}, we defined our measure as the number of rule applications left. For \crule{6}, we measured the "distance until full S" for each concept description in BC. This measure relies on the fact that for any concept description $C \in \BC$, $S(C)$ is a subset of $\BC \cup \{\bot\}$. Since $\BC \cup \{\bot\}$ is finite and the contents of $S$ are monotonically increasing, it gives us a proper measure. Unlike normalization, each rule application reduced the measure, so we didn't need to change the rules.

\subsection{Subsumption Checking}
Our subsumption checker function ties individual steps together and then checks the two conditions described in \baa Listing \ref{check-subsumption} displays the final subsumption checking algorithm between two concept descriptions.

\begin{listing}[tb]%
\caption{Subsumption checking algorithm.}
\label{check-subsumption}
\begin{lstlisting}
check_subsumption$(\cbox, C, D) :=$ 
  $(\cbox^t, A, B) :=$ transform_kb$(\cbox, C, D)$
  $\cbox^n :=$ normalize$(\cbox^t)$
  $\cbox^{A+} :=$ A_extension$(\cbox^n, A)$
  $cl :=$ classify$(\cbox^{A+})$
  if $B \in cl.S(A)$ or $\bot \in cl.S(A)$:
    return true
  for $i \in \cbox^{A+}.N_I$:
    if $\bot \in cl.S(\{i\})$:
      return true
  return false
\end{lstlisting}
\end{listing}

\subsection{String and Rational Number Domains}
We implemented rational number and string domains described in \baa for our runnable code. In our implementation, we took advantage of the flexibility of representing names in our formalism to reduce the complexity of processing the predicate names with built-in arguments, such as ``equals to 1''. Instead of using strings as predicate names, we defined a type that carries those built-in arguments in them. Using a custom type allowed us to eliminate all the manipulation that may be required if strings or other less expressive types are used.

The original paper \cite{Pushing-the-EL-Envelope} suggests a satisfiability algorithm for each concrete domain they present. The algorithm for rational numbers is described in \cite{PSpace-Reasoning-with-the-Description-Logic-ALCF(D)} and constructs a linear program that has a solution if and only if the predicate expression is satisfiable. The algorithm for strings is described in \cite{NEXPTIME-Complete-Description-Logics-With-Concrete-Domains} and uses directed graphs and a specifically designed subroutine to decide the satisfiability. Due to the high complexity of each algorithm, we decided to implement them as an unverified OCaml program that is being called from our verified implementation. We used the OCaml LP library that interfaces with GNU Linear Programming Kit (GLPK)~\cite{GLPK} rational number satisfiability and the OCamlgraph library to implement directed graph functionality.

\section{Proving Correctness}
As presented in the original paper \cite{Pushing-the-EL-Envelope}, we split the correctness of the algorithm into soundness and completeness. To ensure end-to-end correctness, we tied all the proofs together at the end to obtain the following succinct specification:

\begin{theorem} Correctness

    \noindent
    $\forall\ \cbox, C, D.$

    \noindent
    $\text{\texttt{check\_subsumption}}\, (\cbox, C, D) = \text{\texttt{true}} \ifff C \ci_\cbox D.$
\end{theorem}

\subsection{Soundness}
Our soundness proof followed a similar structure to \baa's pen-and-paper proofs. Required properties were invariant under each rule application, therefore we only needed to prove their invariance. The rest of the soundness proof mechanized smoothly.

\subsection{Completeness}
Completeness proof was more challenging to mechanize than soundness proof. It required us to define a new invariant, fix errors in the original proof, and employ non-constructive reasoning. We will explain our new invariant and the errors in the original proof below. 

\paragraph{Role Trees}
Point 2 of Claim 1 in the original completeness proof involved a property that holds for the final classification but doesn't necessarily hold in every intermediate step. The proof in the original paper uses induction over only some rule applications, which requires us to do inductive proof over groups of rule applications instead of each application. This is not an easily mechanizable proof strategy. To mechanize the proof, we defined an invariant that implies the desired property for the final classification but is also preserved during each recursive call.

\begin{listing}
\caption{Definition of a role tree.}
\label{role-tree}
\begin{lstlisting}
Inductive role_tree :=
| $Exists_R\ (r, C, D)$ 
| $Trans_R\ (r, C, D, r_1, t)$
| $Split_R\ (r, C, D, r_1, r_2, E, t_1, t_2)$
\end{lstlisting}
\end{listing}

To define the invariant, we first defined a \emph{role tree}, an inductive structure that holds the information on how a pair of concept descriptions can be included in the classification of a role. Listing \ref{role-tree} displays its formal definition.

Our role tree has three constructors: one for each rule that extends the R component of a classification.
\begin{itemize}
\item $Exists_R$ corresponds to the addition by \textbf{CR3} and forms the leaf nodes.

\item $Trans_R$ corresponds to the addition by \textbf{CR10} where $t$ is the tree that contains the information on how $(C, D)$ can be included in $R(r_1)$.

\item $Split_R$ corresponds to the addition by \textbf{CR11} where $t_1$ and $t_2$ are the trees that contain the information on how $(C, E)$ and $(E, D)$ can be included in $R(r_1)$ and $R(r_2)$, respectively. 
\end{itemize}
  
Even though role trees contained possible historical information, they didn't include what having that history entails. To establish this connection, We defined a function that constructs a proposition from a tree given a knowledge base and a classification. Function recursively traverses a given tree and returns the proposition that asserts that conditions of corresponding rules hold. Listing \ref{tree-to-invariant} shows its definition.

\begin{listing}
\caption{Proposition construction function.}
\label{tree-to-invariant}
\begin{lstlisting}
Fixpoint tree_to_invariant tree :=
match tree with
| $Exists_R\, (r, C, D) \Rightarrow$
    $\exists\, C',\, C' \in S(C) \wedge C' \ci \exists r. D \in \cbox$ 

| $Trans_R\, (r, C, D, r_1, t) \Rightarrow$ 
    tree_to_invariant$\, (t)\, \wedge$
    $(C, D) \in R(r_1) \wedge r_1 \ci r \in \cbox$
    
| $Split_R\, (r, C, D, r_1, r_2, E, t_1, t_2) \Rightarrow$ 
    $E \in \BC\, \wedge$
    tree_to_invariant$\, (t_1)\, \wedge$
    tree_to_invariant$\, (t_2)\, \wedge$
    $(C, E) \in R(r_1) \wedge (E, D) \in R(r_2)\, \wedge$ 
    $r_1 \circ r_2 \ci r \in \cbox$
end.
\end{lstlisting}
\end{listing}

Finally, we defined a predicate that asserts the root of a tree corresponds to a given $(r, C, D)$ tuple. Given these three definitions, our invariant states that for all concept descriptions $C$ and $D$, and role name $r$, if $(C, D)$ is in $R(r)$, then there exists a role tree such that, its root corresponds to $(r, C, D)$ and the derived proposition holds.

Role trees allowed us to mimic doing induction over selected rule applications by doing induction over the trees. Although we didn't need to use this technique again, we believe it is a strong technique to tackle proofs about properties that can be broken temporarily.

\subsection {Errors in Completeness Proofs}
Our mechanization revealed two major errors in pen-and-paper completeness proofs. The first one is the non-transitivity of a relation assumed to be transitive. The second one is the under-specification of a solution for concrete domain predicates. We explain each problem and our solutions below.

\subsubsection{Non-transitivity}
Completeness proofs in \baa heavily rely on a relation we call $\BCm$-equivalence, denoted by $\sim$. 

The relation is defined over the set $\BCm$, which is the set of all concept descriptions in $\BC$ that are reachable from $A$. The relation $C \sim D$ for concept descriptions $C, D \in \BCm$ is defined as either $C$ is equal to $D$, or there exists an individual name $a$ such that $\{a\} \in S(C) \cap S(D)$. Their formal definitions are as follows:
\vspace{6pt}

$\BCm = \{ C \in \BC\ |\ A \reach C\}$

$\sim\ = \{(C, D)\ |\  C = D \vee \exists\ a.\ \{a\} \in S(C) \cap S(D)\}$

\vspace{6pt}
During our formalization, we discovered that $\sim$ is not an equivalence relation because it wasn't transitive. Since the rest of the proof relied on it being an equivalence in multiple places, the completeness proof in the paper wasn't correct. Figure \ref{incomplete-example} shows an example where transitivity doesn't hold. The example contains a transformed and normalized CBox for $\exists r_1.X \ci_\cbox \bot$ query. In the example, $X, \{b\}, \{c\} \in \BCm$, $\{b\} \sim X$, and $X \sim \{c\}$ but it is not true that $\{b\} \sim \{c\}$. Therefore, $\sim$ is not transitive under the old algorithm. It is also important to point out that the example does not demonstrate that the old algorithm is incomplete.

\begin{figure}
\noindent
\textbf{Knowledge base}

$N_C = \{X, A, B\}\quad N_R = \{r_1\}\quad N_I = \{b, c\}$

$\cbox = \{X \sqsubseteq \{b\}, X \sqsubseteq \{c\}, A \sqsubseteq \exists r_1.X\}$
\vspace{6pt}

\noindent
\textbf{Classification result}

$S(X) = \{\top, X, \{b\}, \{c\}\}\quad S(A) = \{\top, A\}$

$S(\{b\}) = \{\top, \{b\}\}\quad S(\{c\}) = \{\top, \{c\}\}$

$R(r_1) = \{(A, X)\}$

\caption{Example CBox where $\sim$ is not transitive.}
\label{incomplete-example}
\end{figure}

Transitivity of the relation relies on the following two properties:

\begin{enumerate}
    \item For all models $\model$ of CBox $\cbox$ and individuals $i$ and concept descriptions $C$ and $D$, if $i$ is in $C^\model$ and $D$ is in $S(C)$, then $i$ is in $D^\model$.
    \item For all models $\model$ of CBox $\cbox$ and concept descriptions $C$ and $D$, if $C \reach D$ and $C^\model$ is nonempty, then $D^\model$ is nonempty.
\end{enumerate}

The problem stemmed from the fact that in some models of $\cbox$, $A^\model$ can be empty. We believe authors implicitly assumed that they could prove the completeness by reasoning about the models where $A^\model$ is nonempty since subsumption holds trivially in the models where $A^\model$ is empty. They used this assumption to prove that 

$$\forall\ C, D \in \BCm.\ C \sim D \rightarrow S(C) = S(D)$$

which doesn't hold if a model exists where $A^\model$ is empty. Since classifications and models are ``living in separate levels,'' the statement cannot be qualified to hold only under the nonempty models, which makes the statement false.

We fixed this error by introducing $A$-extensions that enforce nonemptiness on the classification level. The models of an $A$-extension of a CBox are precisely the ones where $A^\model$ is nonempty. By restricting the models to nonempty ones, we recovered the transitivity of the relation, which fixed the major flaw in the proof of completeness. The question of whether the original algorithm is complete remains open.

\subsubsection{Under-specification}
We identified another error in the statement of a lemma, originally called Claim 2, used in constructing a counterexample model. Lemma stated that for each concrete domain $dm$ and an equivalence class $[C]$ with respect to $\sim$, there exists a solution $\delta([C], dm)$ that only satisfies the predicate tuples in $S(C)$. Below is the original lemma from \baa

\vspace{6pt}
\textbf{Claim 2.} For each $C \in \BCm$ and each concrete domain $dm$, we can find a solution $\delta([C], dm)$ for $\text{\textbf{con}}_{dm}(S(C))$ such that, for all concepts $D \in \BC$ of the form $p(f_1,\cdots,f_k)$ with $p \in \mathcal{P}^{dm}$, we have $\delta([C], dm) \models D$ iff $D \in S(C)$, where $\text{\textbf{con}}_{dm}(S(C))$ denotes the conjunction of all the predicates in $S(C)$ that belongs to $dm$. \footnote{
We modified the notation to make it consistent with our notation. Also, the original statement contained $\delta([C], t) \models D$ for an unknown variable $t$, which we believe is a typo. 
}

\vspace{6pt}
This solution defines the interpretation of feature names in the counterexample. The behavior of solutions from Claim 2 is only restricted in terms of the domain they are being selected for. Therefore, they may satisfy some predicate tuples not in $S(C)$ from other domains. However, proof of Claim 3 relies on the fact that this solution doesn't satisfy any other predicate not in $S(C)$, regardless of whether the predicate belongs to the chosen domain. 

We fixed this error by reordering the quantification and further restricting the behavior of the solution. The definition of our lemma is

\begin{lemma} Solution existence

\noindent
$\exists\ \delta.\ \forall\ C \in \BCm, dm, p \in \mathcal{P}^{dm}, f_1, \cdots, f_n.$

$(\delta([C]) \models p(f_1, \cdots, f_n) \ifff p(f_1, \cdots, f_n) \in S(C))\ \wedge$

$(\forall\ f.\ (\exists\ e.\ \delta([C], f) = e) \ifff$

$\quad \exists\ dm, p \in \mathcal{P}^{dm}, f_1, \cdots, f_n.$

$\quad p(f_1, \cdots, f_n) \in S(C) \wedge f \in (f_1, \cdots, f_n))$
\end{lemma}

Our lemma states that there exists a function $\delta$ from $\sim$-equivalence classes to solutions, i.e., partial functions from feature names to concrete domain elements, such that, for all concept descriptions $C$ in $\BCm$, all concrete domains and all predicate expressions, $\delta([C])$ 
validates the expression if and only if it is in $S(C)$. Lemma also restricts the feature names with elements assigned to them to the ones that appear in $S(C)$. These two combined are strong enough to prove Claim 3.

\section{Extraction}
We used Coq's extraction procedure to obtain a runnable OCaml code. We extracted certain Coq types to native OCaml types for better run-time performance. More precisely, we extracted natural numbers and integers to Ocaml \texttt{int} type, rational numbers to zarith library's \texttt{Q} type, and strings to native Ocaml \texttt{string} type. This choice introduces possible unsoundness to the runnable code due to possible integer overflows since original Coq types are unbounded and cannot overflow. We believe that this is a reasonable tradeoff since 64-bit integers are very unlikely to overflow in practice. Using Coq's default extraction settings to obtain slower but non-overflowing implementation is still possible. We obtained the custom extraction code for these types from mlvoqc library.~\cite{A-verified-optimizer-for-Quantum-circuits}

\subsection{Trusted Computing Base}
Our trusted computing base consists of Coq Proof Assistant, OCaml run-time and libraries, GLPK linear program solver, and our implementations of satisfiability and implication checkers for concrete domains.

\section{Evaluation}
We evaluated our work in two axes. We measured the required proof effort by comparing the number of lines of proof per line of implementation. We evaluated our extracted artifact for runtime performance on randomly generated knowledge bases of various sizes.

\subsection{Proof Effort}
We measured proof effort by lines of code and the number of definitions and theorems. Our 165 definitions consist of 1511 cloc. Our 387 theorems consist of 16977 cloc, giving us 11.3x proof overhead.

\subsection{Performance Evaluation}
To evaluate the performance of our implementation, we measured run times on randomly generated knowledge bases with different sizes and structures. We designed our test cases to measure the impact of concept and role inclusion axioms separately. All knowledge bases had 20 concept names, 10 role names, and 20 individual names. For concept inclusion tests, we kept the number of role inclusion axioms at 10 and the maximum length of the left side at 2. For role inclusion tests, we set the number of concept inclusions to 20 and the maximum length of the left side of role inclusions to 5. All the tests are run on AMD Ryzen 5 2.20 GHz CPU with 16 GB of RAM. The following table shows sample numbers where each row corresponds to a constraint type, and each column corresponds to the number of constraints of that type. 

\begin{center}
\begin{tabular}{ | c | c | c | c | }
\hline
     \# of constraints & 20 & 30 & 40 \\
    \hline
    Concept Inclusions & 4s  & 278s & 192s \\
    \hline
    Role Inclusions & 463s & 6s & 2732s \\
    \hline
\end{tabular}
\end{center}

Our results show a very high variance between different tests, indicating that, as expected, the structure of a knowledge base has more impact than its size when it comes to reasoner performance.

\section{Conclusion and Future Directions}

The Semantic Web, over the past three decades, helped enhance a wide range of applications. Its core principle of providing a semantic layer for data allows for meaningful deductions and explanations, setting it apart from sub-symbolic technologies like neural networks. However, the varying sophistication of SW systems' reasoning abilities highlights the need for reliable reasoning algorithms, particularly in the context of the growing interest in Neurosymbolic AI and Generative AI approaches that seek to utilize Semantic Web layers. The critical importance of ensuring the correctness of reasoner results cannot be overstated, especially as AI systems are increasingly used in sensitive and high-stakes domains. 

In an attempt to address this gap, our work lays the foundation for formally verified reasoners for Neurosymbolic AI applications. We believe that, in the cases of high-stakes application domains, such as health, finance, and security, the benefit of increased trustworthiness outweighs the increased development effort. As AI continues to enter different and critical domains, ensuring the correctness of reasoning processes through formal verification might become essential in preventing errors and fostering trust in AI systems.

As for future directions, one immediate step would be incorporating domain and range restriction into \ELpp to increase its expressive power. This integration is theoretically described in \cite{Pushing-the-EL-Envelope-Further}. To implement the increase in expressive power, we would be required to modify the language and add new functionality to eliminate domain and range restrictions. In relation to this approach, we might work towards implementing reasoners for other OWL profiles since OWL EL's expressive power is limited. 

Another future direction would be improving the implementation performance. Our work faithfully replicated the Baader et al.'s algorithm. During this process, we observed many places where the implementation can be modified for better run-time performance. For example, the algorithm can traverse the knowledge base and group the constraints based on which rule they may fit. This approach would eliminate the overhead of searching for a candidate one at a time, repeat testing of unviable candidates, and enable parallel application of the rules.

\bibliography{aaai25}

\begin{thebibliography}{32}
\providecommand{\natexlab}[1]{#1}

\bibitem[{Baader, Brandt, and Lutz(2005)}]{Pushing-the-EL-Envelope}
Baader, F.; Brandt, S.; and Lutz, C. 2005.
\newblock Pushing the EL Envelope.
\newblock In \emph{Proceedings of the 19th International Joint Conference on Artificial Intelligence}, IJCAI'05, 364–369. San Francisco, CA, USA: Morgan Kaufmann Publishers Inc.

\bibitem[{{Baader}, {Brandt}, and {Lutz}(2008)}]{Pushing-the-EL-Envelope-Further}
{Baader}, F.; {Brandt}, S.; and {Lutz}, C. 2008.
\newblock Pushing the EL Envelope Further.
\newblock In {Clark}, K.; and {Patel-Schneider}, P.~F., eds., \emph{In Proceedings of the OWLED 2008 DC Workshop on OWL: Experiences and Directions}.

\bibitem[{Baader, Koopmann, and Tinelli(2020)}]{First-Results-on-How-to-Certify-Subsumptions-Computed-by-the-EL-Reasoner-ELK-Using-the-Logical-Framework-with-Side-Conditions}
Baader, F.; Koopmann, P.; and Tinelli, C. 2020.
\newblock First Results on How to Certify Subsumptions Computed by the EL Reasoner ELK Using the Logical Framework with Side Conditions.
\newblock \emph{Description Logics}.

\bibitem[{Bereczky et~al.(2022)Bereczky, Chen, Horp{\'{a}}csi, Mizsei, Pe{\~{n}}a, and Tusil}]{Mechanizing-Matching-Logic-in-Coq}
Bereczky, P.; Chen, X.; Horp{\'{a}}csi, D.; Mizsei, T.~B.; Pe{\~{n}}a, L.; and Tusil, J. 2022.
\newblock Mechanizing Matching Logic in Coq.
\newblock In Rusu, V., ed., \emph{Proceedings of the Sixth Working Formal Methods Symposium, {FROM} 2022, "Al. I. Cuza University", Iasi, Romania, 19-20 September, 2022}, volume 369 of \emph{{EPTCS}}, 17--36.

\bibitem[{Bischof et~al.(2018)Bischof, Schenner, Steyskal, and Taupe}]{bischof2018integrating}
Bischof, S.; Schenner, G.; Steyskal, S.; and Taupe, R. 2018.
\newblock Integrating Semantic Web Technologies and ASP for Product Configuration.
\newblock In \emph{ConfWS}, 53--60.

\bibitem[{Colucci, Donini, and Di~Sciascio(2024)}]{colucci2024review}
Colucci, S.; Donini, F.~M.; and Di~Sciascio, E. 2024.
\newblock A review of reasoning characteristics of RDF-based Semantic Web systems.
\newblock \emph{Wiley Interdisciplinary Reviews: Data Mining and Knowledge Discovery}, e1537.

\bibitem[{de~Almeida~Borges(2022)}]{Towards-a-Coq-formalization-of-a-quantified-modal-logic}
de~Almeida~Borges, A. 2022.
\newblock Towards a Coq formalization of a quantified modal logic.
\newblock arXiv:2206.03358.

\bibitem[{{Free Software Foundation}(2024)}]{GLPK}
{Free Software Foundation}. 2024.
\newblock GNU Linear Programming Kit.

\bibitem[{Frumin(2022)}]{Semantic-cut-elimination-for-the-logic-of-bunched-implications}
Frumin, D. 2022.
\newblock Semantic cut elimination for the logic of bunched implications, formalized in Coq.
\newblock In \emph{Proceedings of the 11th ACM SIGPLAN International Conference on Certified Programs and Proofs}, CPP 2022, 291–306. New York, NY, USA: Association for Computing Machinery.
\newblock ISBN 9781450391825.

\bibitem[{Glimm et~al.(2014)Glimm, Horrocks, Motik, Stoilos, and Wang}]{glimm2014hermit}
Glimm, B.; Horrocks, I.; Motik, B.; Stoilos, G.; and Wang, Z. 2014.
\newblock HermiT: an OWL 2 reasoner.
\newblock \emph{Journal of automated reasoning}, 53: 245--269.

\bibitem[{Guo and Yu(2023)}]{A-Comprehensive-Formalization-of-Propositional-Logic}
Guo, D.; and Yu, W. 2023.
\newblock A Comprehensive Formalization of Propositional Logic in Coq: Deduction Systems, Meta-Theorems, and Automation Tactics.
\newblock \emph{Mathematics}, 11(11).

\bibitem[{Herbelin, Kim, and Lee(2017)}]{Formalizing-the-meta-theory-of-first-order-predicate-logic}
Herbelin, H.; Kim, S.; and Lee, G. 2017.
\newblock Formalizing the meta-theory of first-order predicate logic.
\newblock \emph{Journal of the korean Mathematical society}, 54(5): 1521--1536.

\bibitem[{Hidalgo-Doblado et~al.(2014)Hidalgo-Doblado, Alonso-Jim\'{e}nez, Borrego-D\'{\i}az, Mart\'{\i}n-Mateos, and Ruiz-Reina}]{Formally-Verified-Tableau-Based-Reasoners-for-a-Description-Logic}
Hidalgo-Doblado, M.~J.; Alonso-Jim\'{e}nez, J.~A.; Borrego-D\'{\i}az, J.; Mart\'{\i}n-Mateos, F.~J.; and Ruiz-Reina, J.~L. 2014.
\newblock Formally Verified Tableau-Based Reasoners for a Description Logic.
\newblock \emph{J. Autom. Reason.}, 52(3): 331–360.

\bibitem[{Hietala et~al.(2021)Hietala, Rand, Hung, Wu, and Hicks}]{A-verified-optimizer-for-Quantum-circuits}
Hietala, K.; Rand, R.; Hung, S.-H.; Wu, X.; and Hicks, M. 2021.
\newblock A verified optimizer for Quantum circuits.
\newblock \emph{Proc. ACM Program. Lang.}, 5(POPL).

\bibitem[{Hitzler et~al.(2022)Hitzler, Eberhart, Ebrahimi, Sarker, and Zhou}]{nai}
Hitzler, P.; Eberhart, A.; Ebrahimi, M.; Sarker, M.~K.; and Zhou, L. 2022.
\newblock {Neuro-symbolic approaches in artificial intelligence}.
\newblock \emph{National Science Review}, 9(6): nwac035.

\bibitem[{Jung et~al.(2018)Jung, Krebbers, Jourdan, Bizjak, Birkedal, and Dreyer}]{Iris-from-the-ground-up}
Jung, R.; Krebbers, R.; Jourdan, J.-H.; Bizjak, A.; Birkedal, L.; and Dreyer, D. 2018.
\newblock Iris from the ground up: A modular foundation for higher-order concurrent separation logic.
\newblock \emph{Journal of Functional Programming}, 28: e20.

\bibitem[{Kazakov, Kr{\"o}tzsch, and Simancik(2012)}]{kazakov2012elk}
Kazakov, Y.; Kr{\"o}tzsch, M.; and Simancik, F. 2012.
\newblock ELK reasoner: architecture and evaluation.
\newblock In \emph{ORE}. Citeseer.

\bibitem[{Lutz(2002)}]{PSpace-Reasoning-with-the-Description-Logic-ALCF(D)}
Lutz, C. 2002.
\newblock PSpace Reasoning with the Description Logic {ALCF(D)}.
\newblock \emph{Logic Journal of the IGPL}, 10(5): 535--568.

\bibitem[{Lutz(2004)}]{NEXPTIME-Complete-Description-Logics-With-Concrete-Domains}
Lutz, C. 2004.
\newblock NEXP TIME-complete description logics with concrete domains.
\newblock \emph{ACM Trans. Comput. Logic}, 5(4): 669–705.

\bibitem[{McGinty(2018)}]{knarm}
McGinty, H.~K. 2018.
\newblock \emph{KNowledge Acquisition and Representation Methodology (KNARM) and Its Applications}.
\newblock Ph.D. thesis, University of Miami.

\bibitem[{Mishra and Kumar(2011)}]{mishra2011semantic}
Mishra, R.~B.; and Kumar, S. 2011.
\newblock Semantic web reasoners and languages.
\newblock \emph{Artificial Intelligence Review}, 35: 339--368.

\bibitem[{Musen(2015)}]{protege}
Musen, M.~A. 2015.
\newblock The prot{\'e}g{\'e} project: a look back and a look forward.
\newblock \emph{AI matters}, 1(4): 4--12.

\bibitem[{Necula et~al.(2018)Necula, P{\u{a}}v{\u{a}}loaia, Str{\^\i}mbei, and Dospinescu}]{necula2018enhancement}
Necula, S.-C.; P{\u{a}}v{\u{a}}loaia, V.-D.; Str{\^\i}mbei, C.; and Dospinescu, O. 2018.
\newblock Enhancement of e-commerce websites with semantic web technologies.
\newblock \emph{Sustainability}, 10(6): 1955.

\bibitem[{Parsia et~al.(2017)Parsia, Matentzoglu, Gonçalves, Glimm, and Steigmiller}]{The-OWL-Reasoner-Evaluation-ORE-2015-Competition-Report}
Parsia, B.; Matentzoglu, N.; Gonçalves, R.~S.; Glimm, B.; and Steigmiller, A. 2017.
\newblock The OWL Reasoner Evaluation (ORE) 2015 Competition Report.
\newblock \emph{Journal of Automated Reasoning}, 59: 455 -- 482.

\bibitem[{Perera(2020)}]{perera2020personalised}
Perera, M. 2020.
\newblock Personalised human device interaction through context aware augmented reality.
\newblock In \emph{Proceedings of the 2020 International Conference on Multimodal Interaction}, 723--727.

\bibitem[{Sirin et~al.(2007)Sirin, Parsia, Grau, Kalyanpur, and Katz}]{pellet}
Sirin, E.; Parsia, B.; Grau, B.~C.; Kalyanpur, A.; and Katz, Y. 2007.
\newblock Pellet: A practical owl-dl reasoner.
\newblock \emph{Journal of Web Semantics}, 5(2): 51--53.

\bibitem[{Smith et~al.(2007)Smith, Ashburner, Rosse, Bard, Bug, Ceusters, Goldberg, Eilbeck, Ireland, Mungall, Leontis, Rocca-Serra, Ruttenberg, Sansone, Scheuermann, Shah, Whetzel, and Lewis}]{obo}
Smith, B.; Ashburner, M.; Rosse, C.; Bard, J. B.~L.; Bug, W.~J.; Ceusters, W.; Goldberg, L.~J.; Eilbeck, K.; Ireland, A.; Mungall, C.~J.; Leontis, N.~B.; Rocca-Serra, P.; Ruttenberg, A.; Sansone, S.-A.; Scheuermann, R.~H.; Shah, N.~H.; Whetzel, P.~L.; and Lewis, S.~E. 2007.
\newblock The OBO Foundry: coordinated evolution of ontologies to support biomedical data integration.
\newblock \emph{Nature Biotechnology}, 25: 1251--1255.

\bibitem[{Steigmiller, Liebig, and Glimm(2014)}]{steigmiller2014konclude}
Steigmiller, A.; Liebig, T.; and Glimm, B. 2014.
\newblock Konclude: system description.
\newblock \emph{Journal of Web Semantics}, 27: 78--85.

\bibitem[{{The Coq Development Team}(2024)}]{coq}
{The Coq Development Team}. 2024.
\newblock \emph{The {Coq} Proof Assistant, version 8.19.2}.

\bibitem[{Tsarkov and Horrocks(2006)}]{fact++}
Tsarkov, D.; and Horrocks, I. 2006.
\newblock FaCT++ description logic reasoner: System description.
\newblock In \emph{International joint conference on automated reasoning}, 292--297. Springer.

\bibitem[{Xavier et~al.(2018)Xavier, Olarte, Reis, and Nigam}]{Mechanizing-Focused-Linear-Logic-in-Coq}
Xavier, B.; Olarte, C.; Reis, G.; and Nigam, V. 2018.
\newblock Mechanizing Focused Linear Logic in Coq.
\newblock \emph{Electronic Notes in Theoretical Computer Science}, 338: 219--236.
\newblock The 12th Workshop on Logical and Semantic Frameworks, with Applications (LSFA 2017).

\bibitem[{Zanarini, Luna, and Sierra(2012)}]{Alternating-Time-Temporal-Logic-in-the-Calculus-of-CoInductive-Constructions}
Zanarini, D.; Luna, C.; and Sierra, L. 2012.
\newblock Alternating-Time Temporal Logic in the Calculus of (Co)Inductive Constructions.
\newblock In Gheyi, R.; and Naumann, D., eds., \emph{Formal Methods: Foundations and Applications}, 210--225. Berlin, Heidelberg: Springer Berlin Heidelberg.
\newblock ISBN 978-3-642-33296-8.

\end{thebibliography}

\end{document}